\documentclass[twocolumn]{aastex631}
\usepackage{amsmath}
\begin{document}

\title{Tilted Dark Halos are Common, Long-Lived, and can Warp Galactic Disks}

\author[0000-0002-6800-5778]{Jiwon Jesse Han}
\affiliation{Center for Astrophysics $|$ Harvard \& Smithsonian, 60 Garden Street, Cambridge, MA 02138, USA}

\author[0000-0002-6648-7136]{Vadim Semenov}
\affiliation{Center for Astrophysics $|$ Harvard \& Smithsonian, 60 Garden Street, Cambridge, MA 02138, USA}

\author[0000-0002-1590-8551]{Charlie Conroy}
\affiliation{Center for Astrophysics $|$ Harvard \& Smithsonian, 60 Garden Street, Cambridge, MA 02138, USA}

\author{Lars Hernquist}
\affiliation{Center for Astrophysics $|$ Harvard \& Smithsonian, 60 Garden Street, Cambridge, MA 02138, USA}

\begin{abstract}

In the $\Lambda$-CDM paradigm, the dark halo governs the gravitational potential within which a galaxy can form and evolve. In this Letter we show that the present-day inner ($r<50\text{ kpc}$) dark halo can be significantly misaligned with the stellar disk. To this end, we use the TNG50 run from the cosmological magneto-hydrodynamic IllustrisTNG simulation suite. Such ``tilted'' dark halos can arise from a variety of processes including major mergers, massive fly-bys, or interactions with satellite companions. Furthermore, we show that tilted dark halos: (1) are well traced by tilted stellar halos, (2) can maintain their tilt for $>$ 5 Gyr in isolated evolution, and (3) can generate warps in the outer disks that are stable over many Gyr. A tilted dark halo holds clues to important events in the formation history of a galaxy, and could help explain the abundance of warped disks in galaxy observations, including the Milky Way.

\end{abstract}

\section{Introduction} \label{sec:intro}

In a $\Lambda$-CDM universe, dark matter collapses into halos \citep{white78, davis85,rubin85, frenk85} that create the gravitational wells within which galaxies form \citep{blumenthal84}. While the spherically averaged profile of dark halos in $N$-body simulations is roughly universal \citep{navarro97}, it was realized early on that dark halos are likely aspherical in nature \citep{Frenk88, dubinski91}. On the other hand, luminous matter affects the shapes of dark halos, preferentially causing the inner halo to be more spherical \citep[][]{gnedin04, kazantzidis04, gustafsson06, duffy10}. Furthermore, hierarchically formed halos comprise a wealth of substructure that ranges from intact to completely relaxed \citep{bullock05}, which also affects the overall shape of the ``smooth'' dark halo \citep{moore04, cooper10}. An interesting regime is the transition from the inner ($r<0.1\text\:r_\text{virial}$) to outer ($r>0.5\:r_\text{virial}$) halo, where the intrinsic asphericity of the dark halo competes with the backreaction from luminous matter, potentially enhanced or erased by major accretion events. 

Connecting the properties of the dark halo to its directly observable counterpart---the stellar halo---is a valuable tool that allows observations to constrain the dark halo. \textit{Gaia} \citep{dr2} has revealed that the stellar halo of the Milky Way is dominated by one major merger event \citep{belokurov18, helmi18}, which may be connected to a global asymmetry in its distribution \citep{iorio19, han22b}. Furthermore, using the H3 Survey \citep{conroy19}, \cite{han22b} find that the stellar halo from Galactocentric radius $r= 5-50\text{ kpc}$ is $\sim25^\circ$ tilted with respect to the plane of the disk.

An intriguing question is whether the dark halo can also be tilted in a similar direction on these scales \citep{han22a}. \cite{shao21} and \cite{emami21} find that the dark halos of Milky Way-like galaxies in large-volume cosmological simulations \texttt{EAGLE} \citep{schaye15} and Illustris TNG50 \citep{pillepich19} generally change orientations beyond $0.1\:r_\text{virial}$, which is around $20-30\text{ kpc}$ for the Galaxy. On the contrary, \cite{orkney23} have analyzed 10 Milky Way-like galaxies in the zoom-in cosmological simulation \texttt{AURIGA} \citep{grand17} and did not find a present-day tilt of the halo with respect to the disk within 50 kpc. 

\begin{figure*}[t]
    \centering
    \includegraphics[width=\textwidth]{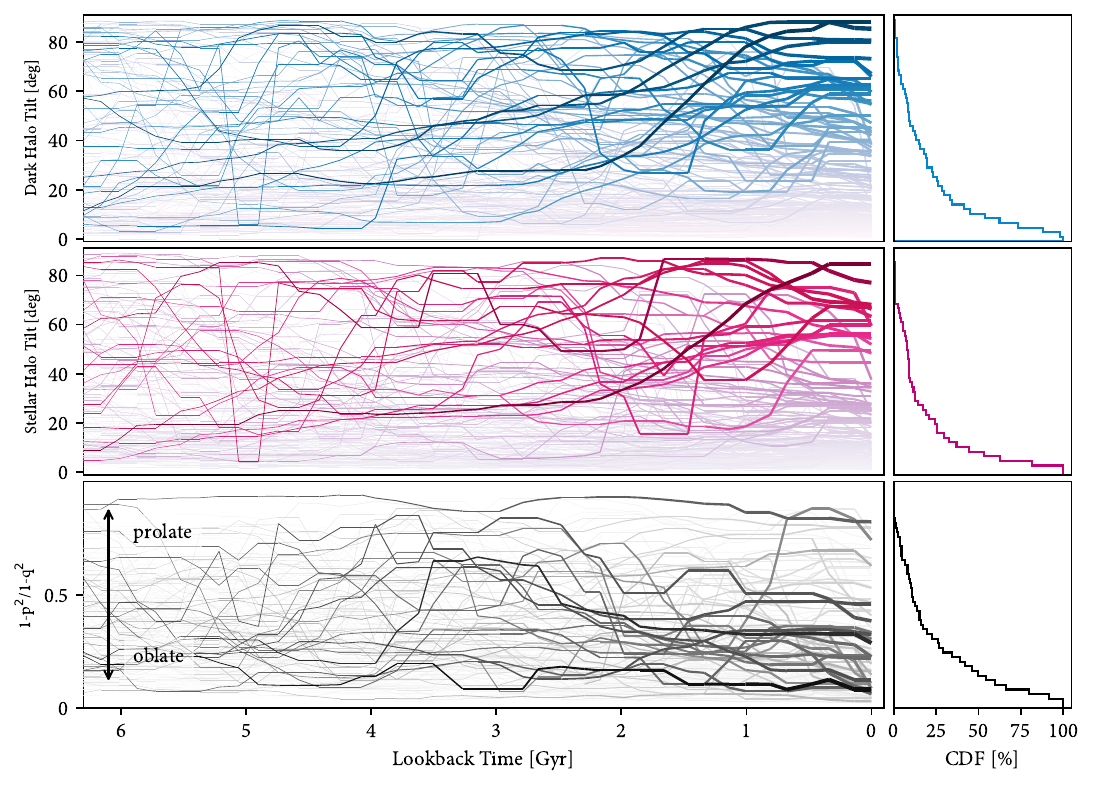}
    \caption{Evolution of the tilt angle of the dark and stellar halo from 6 Gyr ago to present day in TNG50 Milky Way analogs \citep{pillepich23}. Each line represents a single halo. The first and second row panels show the evolution of the tilt angle of the dark matter and stellar halo, and the third row panels show the evolution of the dark halo triaxiality parameter $T\equiv 1-p^2/1-q^2$, where $p$ and $q$ are the major-to-intermediate and major-to-minor axis ratios. We increase the line thickness with time, and color each line according to its final dark halo tilt angle. On the right panel, we plot the cumulative distribution function (CDF) of the tilt angles at the present day, which reveals that around half of TNG50 Milky Way analogs have dark halo tilt angles greater than $10^\circ$. We find a diversity of triaxiality parameters for the tilted halos, ranging from prolate ($T>0.6$) to oblate ($T<0.3$).}
    \label{fig:tree}
\end{figure*}

\begin{figure}[h]
    \centering
    \includegraphics[width=0.5\textwidth]{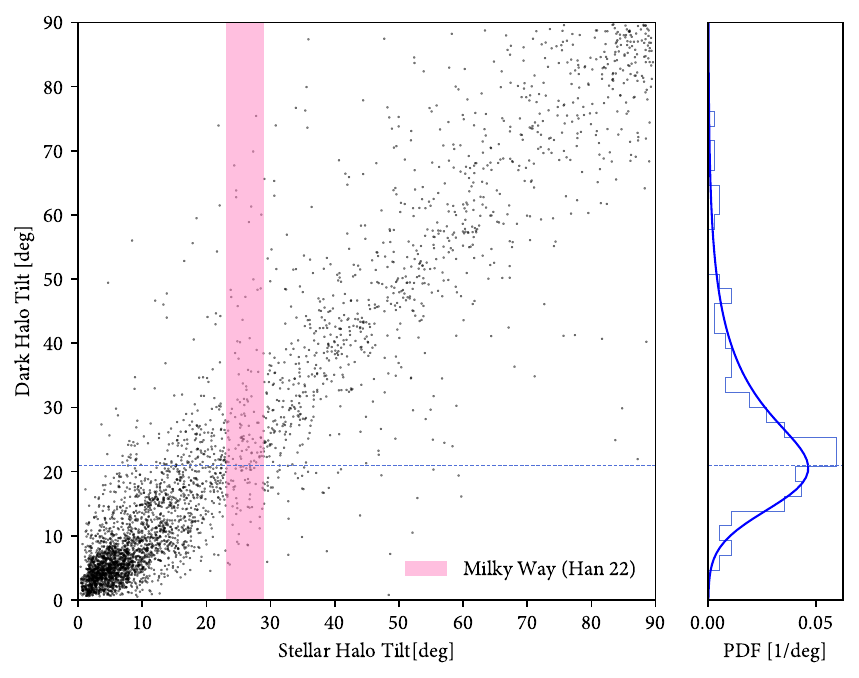}
    \caption{Relationship between the dark halo and the stellar halo tilt angles. Each point shows the dark halo and stellar halo tilt angle at one snapshot, spanning all of the halos and lookback times from Fig \ref{fig:tree}. The pink shaded region marks the measured Galactic stellar halo tilt from \cite{han22b}, and the blue histogram on the right panel shows the marginal distribution of the Milky Way's dark halo. The blue horizontal line shows the most likely value of the Galactic dark halo to be $20^\circ$.}
    \label{fig:scatter}
\end{figure}

In this Letter we address three questions. First, can a present-day tilt of the inner 50 kpc dark halo with respect to the disk \textit{exist}? Second, if those features do exist, are they common and long-lived? Lastly, how do such misalignments affect the stellar disk? To answer these questions, we use the TNG50 run of the cosmological magneto-hydrodynamic simulation suite IllustrisTNG \citep{nelson19,nelson19b,pillepich19}. TNG50 is uniquely suited for our study because it: (1) captures a wide range of galaxy formation scenarios ($\sim\:$200 Milky Way-like halos), (2) incorporates reasonably realistic baryonic physics, and (3) sufficiently resolves the particle dynamics of the halo and the disk.

We organize the Letter as follows. In Section \ref{methods}, we desribe the TNG50 sample of galaxies that we use, then outline the method to analyze their tilt angles. In Section \ref{results}, we identify an archetypal galaxy and closely follow the evolution of its dark matter halo, stellar halo, and stellar disk. Finally, in Section \ref{discussion}, we summarize the results and discuss their implications in a broader context.

\section{Milky Way Analogs in TNG50}\label{methods}

In this Section, we describe the sample of galaxies in TNG50 that we use, and how we define and measure the tilt angles of galaxies over their formation histories. We then present the distribution of dark halo tilt angles in our sample, and relate those measurements to the stellar halo.

\subsection{Sample and Methods}

TNG50 \citep{pillepich19, nelson19,nelson19b} is a cosmological magneto-hydrodynamical simulation run with \texttt{AREPO} \citep{springel10, weinberger10} using the fiducial TNG galaxy formation model \citep{weinberger17, pillepich18} and \cite{Planck16} cosmological parameters. Like the other simulations in IllustrisTNG \citep{pillepich18b,springel18,nelson18,marinacci18,naiman18}, TNG50 follows the evolution of cold dark matter (CDM), stars, gas, magnetic fields, and supermassive black holes from redshift $z=127$ to $z=0$. In the case of TNG50, the volume modeled is ${51.7 \text{ Mpc}}^3$, the mass resolution of the CDM particles is $4.5\times10^{5} M_\odot$, and that of the stellar particles is within a factor of 2 from the target baryonic mass resolution of $8.5\times10^4M_\odot$. Combined with this resolution and the cosmological volume, TNG50 captures a wide range of galaxy formation processes from massive galaxy clusters to isolated dwarf galaxies. At $z=0$, there are $\sim900,000$ halos and subhalos with gravitationally bound mass greater than $10^8M_\odot$. Details of the simulation can be found in \cite{nelson19,nelson19b, pillepich19}.

Using TNG50, \cite{pillepich23} identify Milky Way and M31-like galaxies in the simulation. For this study, we use their ``observable-based selection,'' which is the subset of three criteria: (1) $10.5<\log_{10}(M_*/M_\odot)<2\times10^{12}$, (2) no massive galaxy within $500\text{ kpc}$ and $M_{\text{host}, 200c}<10^{13}M_\odot$, and (3) disky galaxies \citep[based on][and visual inspection]{pillepich19}. This selection yields 198 galaxies and their host halos.

For each halo, we define the ``tilt angle'' to be a misalignment of the stellar disk and the inner dark halo, and calculate this quantity using the following method. We first select dark matter particles with $r \in(10 \text{ kpc},50\text{ kpc})$. From the position and mass of each particle, we calculate the moment of inertia tensor. By solving for the eigenvector-eigenvalue pairs of the tensor, we find the three principal axes of rotation and their respective moments of inertia. The major axis has the minimum moment of inertia, and the minor axis has the maximum moment of inertia. The moments of inertia $I_i$, $i\in\{a,b,c\}$, can be related to the ``length'' $r_i$ of the principal axes as the following:
\begin{equation}
    I_{i} \propto r_{j}^2 + r_{k}^2 \rightarrow r_{i}^2 \propto \frac{-I_i + I_j + I_k}{2}
\end{equation}
Using this relation, we can compute the major-to-intermediate and major-to-minor axes ratios as follows:
\begin{equation}
    1:\frac{r_b}{r_a}:\frac{r_c}{r_a} = 1:\sqrt{\frac{I_a-I_b+I_c}{-I_a+I_b+I_c}}:\sqrt{\frac{I_a+I_b-I_c}{-I_a+I_b+I_c}}
\end{equation}

Some studies calculate the ``reduced'' moment of inertia to down-weight the outer halo particles \citep[e.g.,][]{allgood06, Vera-Ciro11, schneider12, emami21}. In this study, we are limiting the analysis to particles with $10\text{ kpc} < r < 50\text{ kpc}$, and do not downweight the particles. The canonical moment of inertia is sufficient to capture the misalignment of the halo and the disk, and also allows for an easier interpretation of the result. Furthermore, the tilt angle is insensitive to the actual values of the moments of inertia; it is only sensitive to the direction of the principal axes.

Once we obtain the principal axes, we measure the tilt angle as the angle between the \textit{minor} axis and the $Z$-axis, which is set by the total angular momentum of the stellar disk. There are two motivations to use the minor axis as opposed to the major axis. First, many TNG50 galaxies show an oblate inner halo (see last row of Fig. \ref{fig:tree}), meaning that the major and intermediate axes are roughly degenerate with each other. This is consistent with previous studies \citep[e.g.,][]{kazantzidis04,shao21}. In this scenario, the degeneracy between the major/intermediate axes (and the corresponding freedom of azimuthal rotation) causes their angle with respect to the $Z$-axis to be an unstable measurement. On the other hand, the minor axis is almost never degenerate, making its angle with the $Z$-axis a much more stable measurement. Secondly, when the minor axis is aligned with the $Z$-axis, the halo is stable against perturbations to the gravity of the disk, and when the minor axis is $90^{\circ}$ to the $Z$-axis, the halo is unstable. This analogy yields an intuitive interpretation of the tilt angle. We note that the stellar halo tends to be more triaxial than the dark halo in TNG50, which means that all three axes are nondegnerate and one can use either the major or minor axis to measure the tilt angle.

\vspace{15mm}

\subsection{Diversity of Tilt Angles}

In Figure \ref{fig:tree} we show the evolution of the tilt angle of all 198 Milky Way analogs from 6 Gyr ago to the present day. Each colored line represents an individual halo. The top panels show the tilt angle measured for the dark halo, and the middle panels show that of the stellar halo. The bottom panels show the dark halo triaxiality parameter $T\equiv 1-p^2/1-q^2$, where $p$ and $q$ are the major-to-intermediate and major-to-minor axis ratios. High values of this parameter ($T>0.6$) indicate prolate halos, while low values ($T<0.3$) indicate oblate halos. For all rows, we show a cumulative distribution function (CDF) on the right panel that encapsulates the population trend. For example, 50\% of dark halos have tilt angles greater than $10^\circ$, 25\% of dark halos have tilt angles greater than $20^\circ$, and 15\% of dark halos have tilt angles greater than $40^\circ$. This figure demonstrates that the majority of halos, for the majority of their lifetimes, have tilted dark halos. Furthermore, tilted dark halos show a broad range of shapes ranging from prolate to oblate.

In Figure \ref{fig:scatter}, we plot the dark halo tilt angle against the stellar halo tilt angle at each snapshot, for all galaxies in Figure \ref{fig:tree}. The relationship is strikingly linear. The pink region marks the posterior probability of the Milky Way's stellar halo tilt angle from \cite{han22b}, and the consequent marginal probability distribution of the dark halo is plotted in the right panel. We overplot a fitted exponentially modified Gaussian distribution in blue. From this Figure, we can infer that the tilt of the dark halo is most likely greather than $20^\circ$ for the Milky Way at the present day.

\begin{figure}[t]
    \centering
    \includegraphics[width=0.5\textwidth]{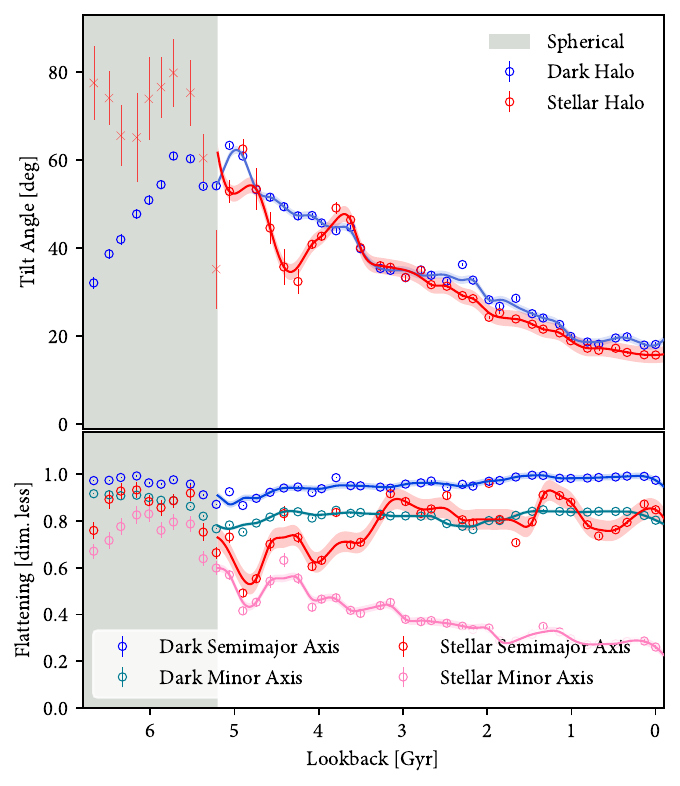}
    \caption{Evolution of Halo 533060 after a 10:1 merger at 7 Gyr. In the top panel, we plot the tilt angle of the dark (stellar) halo in blue (red) dots with $1\sigma$ error bars estimated from jack-knifing. X-marks indicate where the tilt angle uncertainties are larger than $10^\circ$. The colored shaded lines are cubic spline interpolations of the data, sampled from their statistical errors. In the bottom panel, we show the major-to-intermediate and the major-to-minor ratios of the dark/stellar halo, which we refer to as the ``flattening parameters.'' We grey out the regions in which both flattening parameters are close to 1 (i.e. roughly spherical), because the tilt angle is not well defined in this region. Once the dark (stellar) halo settles into an oblate (triaxial) shape at $\sim$5 Gyr, the tilt angle steadily decreases from $50^\circ$ to $20^\circ$ at present day.}
    \label{fig:decay}
\end{figure}

\begin{figure*}[t]
    \centering
    \includegraphics[width=\textwidth]{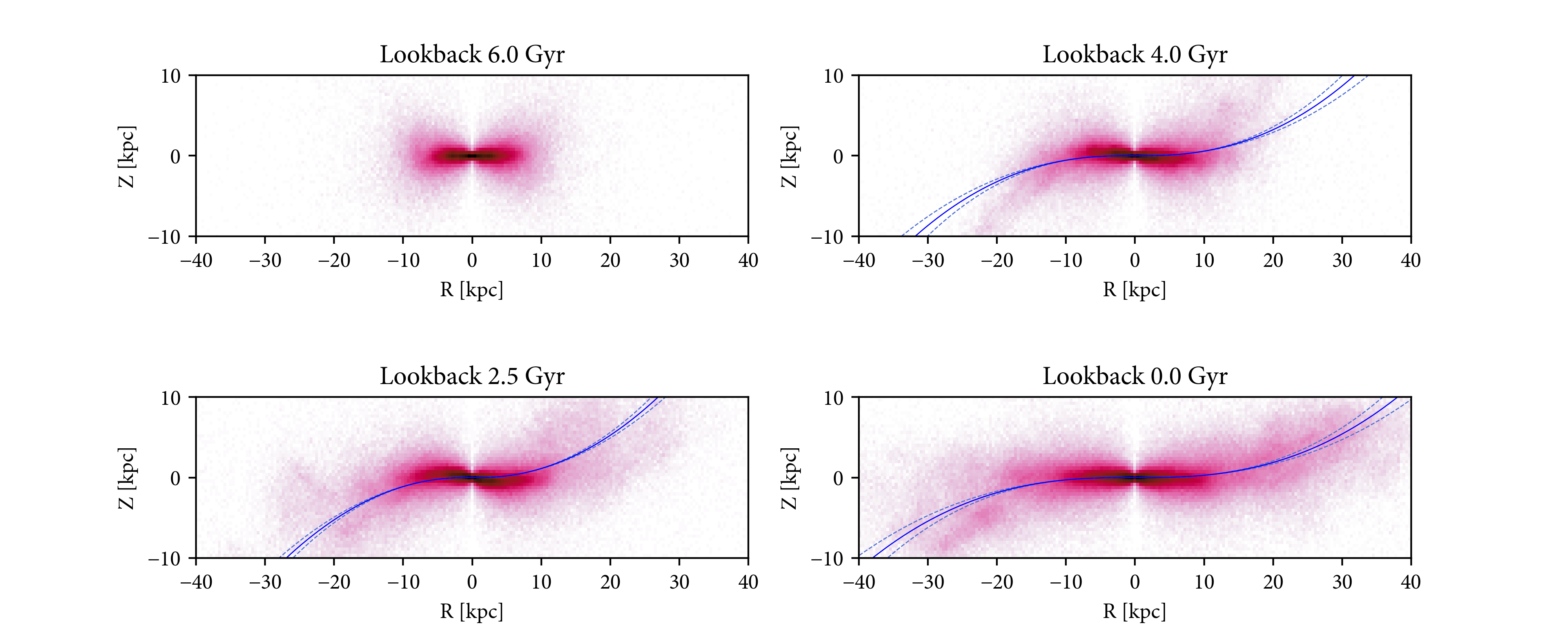}
    \caption{Stellar disk of Halo 533060 plotted at four snapshots. The tilt of the dark halo settles at around 5 Gyr (see Fig \ref{fig:decay}). $Z$ denotes the vetical height of the star, and $R$ denotes the cylindrical radius with positive (negative) sign where the angles are $\pm90^{\circ}$ within the positive (negative) vertical extremum of the warp. We overplot an analytic fit to the warp (power-law in radius and sinusoid in azimuth) in blue, with dotted regions marking $1\sigma$ uncertainties of the fit. A strong warp persists over $>4\text{ Gyr}$.}
    \label{fig:fourwarp}
\end{figure*}

\section{An Archetypal Galaxy}\label{results}

In this Section, we present a case-study of a galaxy with a tilted dark halo and a warped disk. Halo 533060 has a present-day total mass of $8\times10^{11}M_\odot$, and it experiences a 10:1 merger $\sim7\text{ Gyr}$ ago that induces a strong misalignment in the dark halo and the disk. This halo does not undergo any other significant perturbations (e.g., mergers, fly-bys, or massive companions) afterwards, allowing us to study the secular evolution of the tilted dark halo. In the following, we explore the evolution of the halo tilt angle and the response of the disk in Halo 533060.

\subsection{A Tilted Halo}
In Figure \ref{fig:decay} we show the evolution of the tilt and shape of Halo 533060. In the top panel, we plot the tilt angle of the dark (stellar) halo in blue (red) circles, and indicate points with $>5^\circ$ statistical uncertainties with x-marks. The colored shaded regions show a cubic spline interpolation of the points and the $1\sigma$ variance in the interpolated curve. In the bottom panel, we plot the  major-to-intermediate and the major-to-minor axes ratios for the dark/stellar halo, which refer to as the ``flattening parameters.'' We grey out the regions in which both parameters are close to one, indicating that the halo is roughly spherical and the tilt angle is not well defined. For Halo 533060, this region corresponds to the immediate aftermath of the merger. Once the merger is complete, the debris eventually settles and the dark (stellar) halo takes on an oblate (triaxial) shape that is $\sim50^\circ$ misaligned with the disk. Consistent with what we find in Figure \ref{fig:scatter}, the stellar halo tilt angle closely follows that of the dark halo. In subsequent isolated evolution, the dark halo tilt angle declines to $20^\circ$ in 5 Gyr, likely due to a combination of processes such as dynamical friction, phase mixing, and torque exerted by the disk. Furthermore, we see the dark halo shape remains roughly constant, indicating that it rotates as a solid body. This is consistent with previous work \citep[e.g.,][]{bailin04, perryman14}. At the present day, the tilt angle of the stellar halo matches that of the dark halo.

\begin{figure}[t]
    \centering
    \includegraphics[width=0.5\textwidth]{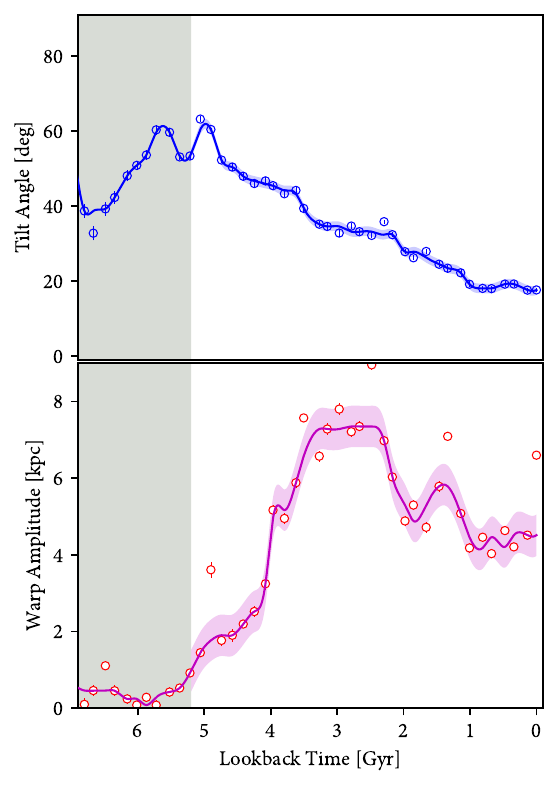}
    \caption{Correspondence between the warp amplitude and the dark halo tilt. On the top panel we plot the evolution of the dark halo tilt as in Fig \ref{fig:decay}. On the bottom panel we plot the amplitude of the disk warp measured at twice the scale radius. We grey out the same region as Fig \ref{fig:decay} where the halo is roughly spherical and the tilt angle is not well defined. When the halo is spherical, the warp amplitude is zero. Once the halo settles to an oblate shape and a tilt angle of $50^\circ$, the warp amplitude sharply increases over 1 Gyr. There are no satellite galaxies or equivalent perturbers present at this epoch. In subsequent evolution, the warp amplitude stays stable over 4 Gyr as the tilt angle of the dark halo slowly declines.}
    \label{fig:warpevolve}
\end{figure}

\subsection{A Warped Disk}

In Figure \ref{fig:fourwarp} we show the stellar disk of Halo 533060 at four representative snapshots. We plot stars in a cylindrical projection, where $R$ is defined to be positive (negative) where the azimuthal angle is $\pm90^\circ$ of the maximum (minimum) vertical height of the disk.  At $t_{\text{lookback}}=6\text{ Gyr}$, there is no discernible azimuthal asymmtery of the disk. However, shortly after the tilted halo settles at $t_{\text{lookback}}=4\text{ Gyr}$, we see a warp beginning to develop in the disk. In subsequent snapshots, a strong warp persists in the disk. We fit the warp as a power-law in radius and sinusoid in azimuth, plotted in blue lines. 

In Figure \ref{fig:warpevolve}, we show the time evolution of the amplitude of the warp alongside the time evolution of the tilted dark halo. We define the warp amplitude to be that of the sinusoid measured at twice the scale radius of the disk, which grows with time \citep[see, e.g.,][for evidence of such ``inside-out'' growth in the Galaxy]{frankel19}. The warp amplitude is zero when the halo is roughly spherical, as shown in the grey shaded region. Once the dark halo settles into an oblate shape with a $50^\circ$ tilt, the warp amplitude sharply increases up to $8\text{ kpc}$ at an approximate rate of $4\text{ kpc} / \text{Gyr}$, then slowly declines to $4\text{ kpc}$ as the tilt angle decreases.

In a companion study, Han et al. (in prep) use idealized simulations to show that a tilted dark halo can induce a warp in the galactic disk within a Gyr. Furthermore, there are no massive satellites around Halo 533060 at the time of the onset of the warped disk. All of these pieces of evidence indicate that the warped disk of Halo 533060 is driven by its tilted dark halo.

\section{Summary and Discussion} \label{discussion}

In this Letter, we have analyzed the ``tilt angle''---a misalignment of the inner ($r<50\text{ kpc}$) dark halo and the stellar disk---in the Illustris TNG50 simulation. We find an abundance of Milky Way-like halos that are significantly tilted at the present day, and identify an isolated halo that allows us to study the secular evolution of a merger-induced tilted dark halo. We find that the tilt angle decreases over time, likely due to dynamical friction and phase mixing of the merger debris. The timescale of this decay is long: the tilt angle declines from $50^\circ$ to $20^\circ$ over 5 Gyr. Furthermore, the stellar halo is a good tracer of the underlying dark halo tilt angle over all epochs. Lastly, we find a compelling relationship between the tilted dark halo and a persistent warp in the galactic disk. The warp amplitude increases sharply after the tilt angle reaches its maximum, and subsequently follows the gradual decline of the tilt angle over $\sim4\text{ Gyr}$.

The shape and configuration of the dark halo is primarily set by gravity, and is affected by baryonic feedback physics \citep[e.g.,][]{debattista08,duffy10}. TNG50 implements a wide range of gravitational and baryonic physical processes including stellar feedback, and reproduces several aspects of observed galaxies \citep{pillepich19}. Thus, following our demonstration of the diversity of tilted dark halos in TNG50, we believe that real galactic halos must also exhibit such diversity.

The origin of the warped disks in galaxies including the Milky Way is a longstanding mystery \citep{kerr57,burke57}. \cite{bosma91} find that at least half of spiral galaxies are warped, which suggests that the warping mechanism must be universal and long-lasting. \cite{semczuk20} have shown in TNG100 that 35\% of the warped disks are caused by satellite encounters (half of which lead to a merger). These satellite-driven warps are transient, surviving on average about 1 Gyr. In this study, we have shown that a tilted dark halo can sustain a long-lasting warp ($>$ 4 Gyr), and that half of the Milky Way analogs have tilt angles greater than $10^\circ$. For the Milky Way, Han et al. (in prep.) demonstrate the same process in an idealized simulation. Using a tilted, triaxial dark halo component based on the observed stellar halo---i.e., adopting a ``minimal-tuning'' model for the dark halo---they calculate the orbits of disk particles. Remarkably, the disk develops a stable warp within 1 Gyr at precisely the observed amplitude, direction, and shape of the Galactic disk warp \textit{and} flare \citep{grabelsky87, bosma91, levine06}. Thus, the results from TNG50 and the isolated simulation make a compelling case for tilted dark halos being the primary driver of warped galaxies in the observed universe.

\bibliography{refs}{}
\bibliographystyle{aasjournal}

\end{document}